# Electronic Devices Based on Purified Carbon Nanotubes Grown By High Pressure Decomposition of Carbon Monoxide

Danvers E. Johnston, Mohammad F. Islam, Arjun G. Yodh, and Alan T. Johnson

Department of Physics and Astronomy, University of Pennsylvania, Philadelphia, Pennsylvania 19104
(Submitted: February 7, 2005)

**The excellent properties of transistors, wires, and sensors made from single-walled carbon nanotubes (SWNTs) make them promising candidates for use in advanced nanoelectronic systems.[1] Gas-phase growth procedures such as the high pressure decomposition of carbon monoxide (HiPCO) method[2,3] yield large quantities of small-diameter semiconducting SWNTs, which are ideal for use in nanoelectronic circuits. As-grown HiPCO material, however, commonly contains a large fraction of carbonaceous impurities that degrade properties of SWNT devices.[4] Here we demonstrate a purification, deposition, and fabrication process that yields devices consisting of metallic and semiconducting nanotubes with electronic characteristics vastly superior to those of circuits made from raw HiPCO. Source-drain current measurements on the circuits as a function of temperature and backgate voltage are used to quantify the energy gap of semiconducting nanotubes in a field effect transistor geometry. This work demonstrates significant progress towards the goal of producing complex integrated circuits from bulk-grown SWNT material.**

To date most work on nanotube electronics has relied on SWNTs grown directly onto substrates by chemical vapor deposition (CVD).[5-7] CVD-grown SWNTs have clean sidewalls enabling high-quality electrical contact to metal electrodes,[8,9] and they hold promise for accurate placement in integrated circuits using patterned catalysts. However, the use of such material for integrated devices is made difficult by lack of control over whether individual SWNTs are metallic or semiconducting. Moreover, semiconducting SWNTs grown by CVD are typically large diameter (2 – 4 nm), leading to an energy band gap much less than 1 eV. This makes them incompatible with designs for nanoelectronic logic gates consisting of SWNT transistors with high ON/OFF ratios.

Gas-phase growth procedures such as the HiPCO method routinely yield large quantities of SWNTs with diameters less than 1 nm and are being scaled to industrial quantities. Growth conditions producing semiconducting SWNTs with a narrow distribution of wrapping vector have been reported,[10] and dispersion of HiPCO material as isolated nanotubes has been achieved using amphiphilic molecules,[11-14] enabling schemes for sorting SWNTs by length,[15,16] diameter,[13] or metallicity.[17,18] Finally, dispersed SWNTs have been controllably deposited onto patterned substrates, exploiting specific interactions between the adsorbed amphiphilic molecules and surface monolayers.[19,20]

A major challenge to realizing the potential of HiPCO material for devices is that it usually contains a significant quantity of carbonaceous impurities known to have deleterious effects on the properties of single nanotube devices. Standard acid or oxidation based purification approaches damage SWNTs and sharply degrade their electronic characteristics (Paulson, S., Wadhar, H., Staii, C., & Johnson, A.T. to be published). In addition, one must remove the amphiphilic molecules after depositing SWNTs onto substrates since residual surfactant on the SWNT leads to poor contact to the electrodes.

Details of the new purification process are provided as Supplemental Information. Briefly, as-grown HiPCO material is purified by heating in wet air in the presence of $H_2O_2$, gentle acid treatment, magnetic fractionation,[21] and vacuum annealing. The dominant impurities in as-grown HiPCO are catalyst particles and non-SWNT carbon phases. Thermogravimetric analysis and wide-angle X-ray scattering measurements indicate impurity content is more than 50 wt% in as-grown HiPCO and less than 5 wt% after purification. Based on this measured impurity content and the measured sample mass after purification, the purification process recovers close to 90% of the SWNT content of the HiPCO. The material to be tested (either raw or purified HiPCO) is dispersed in water using sodium dodecyl benzene sulfonate (NaDDBS)[11] and deposited onto degenerately doped oxidized (400 nm SiO$_2$) silicon wafers. Prior to deposition, the SiO$_2$ surface is functionalized with a 3–aminopropyl triethoxysilane (APTS) monolayer, and SWNTs are deposited by briefly dipping the chip in the SWNT-NaDDBS suspension. The sample is rinsed in deionized water, blown dry, and heated in air at 200 °C for 12 h. This last step removes a large fraction of the residual surfactant as evidenced by a systematic ~ 2 nm decrease of the nanotube diameter as measured by atomic force microscopy (AFM). This treatment also vaporizes the APTS monolayer from the bulk of the silicon substrate.

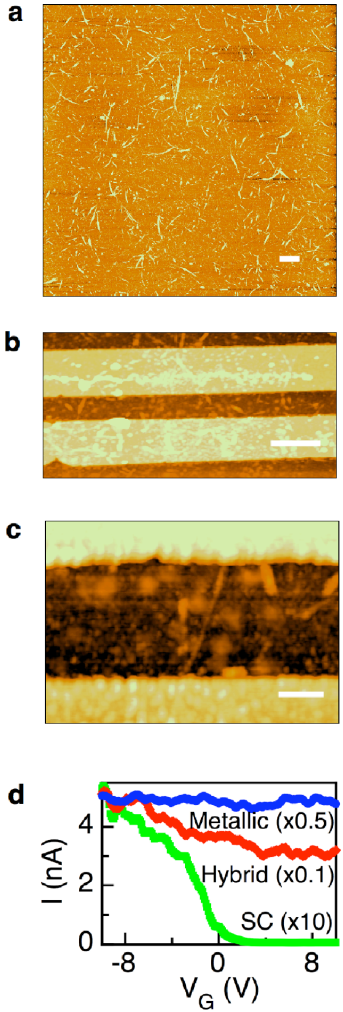

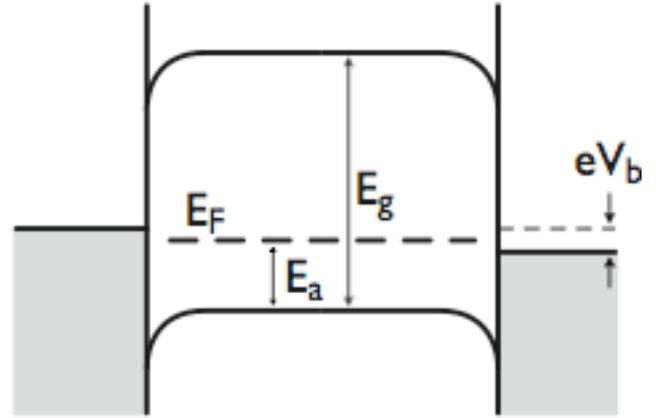

Figure 2 **Model of device energy bands.** Schottky barriers form where metal leads contact a semiconducting SWNT. The Schottky barriers are asymmetric so holes conduct more readily than electrons. Carriers tunnel through the Schottky barriers, so transport is characterized by an activation energy $E_a$ given by the difference between the Fermi energy $E_F$ and the edge of the nearest energy band of the SWNT (here, the valence band).

Figure 1 **AFM images of devices made from purified HiPCO material.** **a** SiO$_2$ surface with individual SWNTs and small bundles after deposition from solution and surfactant removal (scale bar 1 $\mu$m). **b** Cr/Au electrodes contacting SWNT material (scale bar 1 $\mu$m). **c** High resolution scan of a 4-nm diameter bundle with source and drain electrodes along top and bottom (scale bar 200 nm). **d** Three categories of $I(V_g)$ behavior are observed. Bias voltage is 10 mV.

Figure 1a is an AFM image of individual SWNTs and small nanotube bundles after deposition from solution and surfactant removal. Cr/Au source and drain electrodes separated by 400 nm are fabricated with electron beam lithography without alignment followed by thermal evaporation and liftoff (Fig. 1b-c). The electrode density is chosen so ~50% of the electrode pairs conduct, typically contacting one SWNT or one small bundle. The degenerately doped Si is used as a back gate electrode in a field effect transistor (FET) geometry.

The source-drain current $I$ is measured in ambient conditions for different values of the bias voltage $V_b$ and gate voltage $V_g$. The behavior of the $I$–$V_g$ curve at low voltage bias (typical $V_b$ = 10-100 mV) is used to categorize each sample as "metallic" (M), "semiconducting" (SC), or "hybrid" (H). "Metallic" samples have a relatively low source-drain resistance and $I$ shows little or no gate response; we conclude these samples consist of a single metallic SWNT or a bundle where only metallic SWNTs are contacted. "Semiconducting" samples exhibit a high ON/OFF ratio, with very large resistance in the OFF state. We presume conduction occurs through a single semiconducting SWNT or a bundle where current is carried only by semiconducting nanotubes. "Hybrid" samples exhibit a small ON/OFF ratio of roughly 2– 4. We attribute this behavior to conduction by metallic and semiconducting SWNTs in parallel. Figure 1d shows examples of these three observed behaviors.

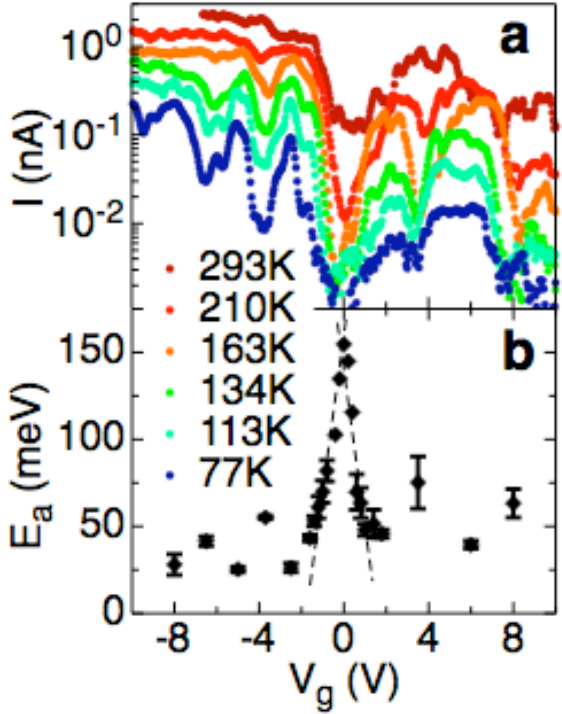

Figure 3 **Current ($I$) – back gate voltage ($V_g$) characteristics for Device I. a** $I(V_g)$ at $V_b = 100$ mV for temperatures 77- 300 K. **b** Thermal activation energy $E_a$ as a function of $V_g$. The peak in $E_a$ corresponds to Fermi energy alignment at midgap. Since the maximum of $E_a$ is 150 meV, the energy gap is found to be 300 meV. The lever arm $\alpha \approx 0.08$ is inferred from the slope of a linear fit to $E_a$ in the gap region. Oscillations in $E_a$ outside the gap region are due to single electron charging.

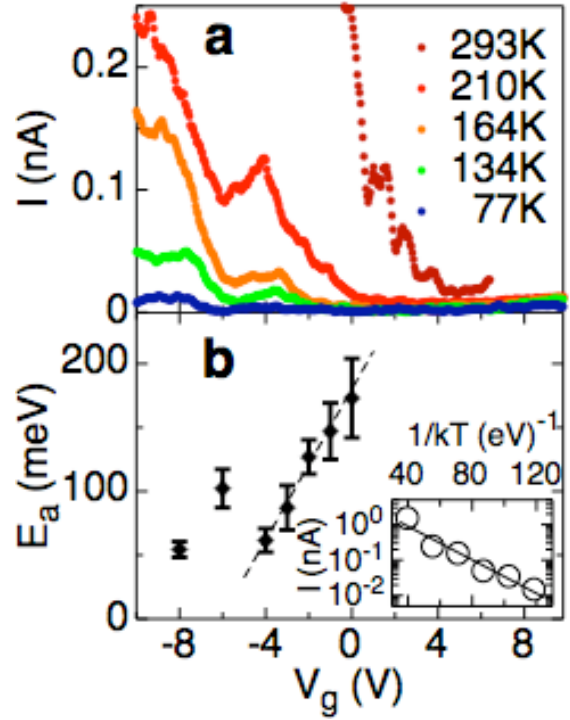

Figure 4 **Current – back gate voltage characteristics for Device II. a** Temperature dependence of $I(V_g)$ with $V_b = 100$ mV. **b** Activation energy $E_a$ as a function of gate voltage. From the maximum value of $E_a$ we find that the energy gap $E_g \geq 400$ mV. The lever arm for this sample is $\alpha \approx 0.03$. Oscillations in $I(V_g)$ and $E_a(V_g)$ for $V_g < -4$ V are due to single electron charging effects. *inset* Arrhenius plot used to find the activation energy for $V_g = -8$ V.

The quality of the purification process was tested by comparing circuits made using as-grown and purified samples from the same HiPCO batch. The fraction of conducting samples was ~25% (4/16 for raw and 7/30 for purified material) for both fabrication runs. Quoted resistance values for SC samples are for the "ON" state ($V_g = -10$ V). M and SC circuits made from raw HiPCO had source-drain resistances near 1 GΩ, while for purified material we measured a median resistance of 4 MΩ for M/H samples; no SC samples were observed in this first trial. Purification thus leads to a decrease in sample resistance by a factor of more than 200. The electrical transport properties of 29 additional samples made from purified material were then measured and classified. Twenty-two samples were M/H with a median resistance of 500 kΩ. Seven samples were SC with a median resistance of 10 MΩ. These should be compared with typical resistances of 15 kΩ and 100 kΩ for M and SC circuits made in our lab with CVD-grown SWNTs. The typical ON/OFF ratio of SC devices was 300, with the highest exceeding 5000. Six of the SC samples exhibited p-type gate behavior similar to FETs made from CVD-grown SWNTs; one SC device had an ambipolar gate response, with both hole and electron conduction (Figure 3a). Supplementary Table I is a complete listing of observed sample resistances.

The observed fraction of SC samples (24%) is consistent with HiPCO material having random chirality (i.e., 2/3 semiconducting SWNTs and 1/3 metallic). If we assume small SWNT bundles (2–4 nm diameter measured by AFM as seen in Figure 1a) show SC behavior only if all of the 3–4 SWNTs on the bundle exterior contacted by the electrodes are semiconducting[22] then we expect 20–30 % of samples to be SC, in satisfactory agreement with the data. However, more single-tube circuits must be measured to precisely quantify the distribution of metallic and semiconducting SWNTs produced by the HiPCO process.

Three sources increase the resistance in SWNT circuits above the quantum limit of $h/4e^2 \cong 6.4\ k\Omega$. Contaminants on the SWNT sidewall increase contact resistance by acting as tunnel barriers at the electrodes or causing poor wetting of the electrode metallization. Schottky barriers form at the contacts to semiconducting (but not metallic) SWNTs, with minimum (tunnel) resistance near $100\ k\Omega$.[23,24] Finally, electron backscattering along the length of the SWNT contributes to resistance. The carrier mean free path for HiPCO is unknown but it can be several micrometers for clean metallic and semiconducting[25] SWNTs grown by CVD.

The high, nearly equal, resistances observed for M and SC devices from as-grown HiPCO indicate that in these samples sidewall contamination is the dominant source of resistance. The new purification process reduces the resistance of both types of samples by a factor of several hundred or more. Despite this improvement, devices from purified HiPCO have resistances significantly larger than those produced from CVD SWNTs. Further experiments will be needed to determine whether this is due to residual contamination that can be removed by an optimized purification process or a larger defect density in purified HiPCO material compared to CVD-grown SWNTs.

Temperature dependent measurements of SC circuits made from purified material are consistent with thermally activated transport, with an activation energy $E_a$ that varies with gate voltage (i.e., $I(V_g,T) \propto e^{-E_a(V_g)/k_B T}$, where $k_B$ is Boltzmann's constant). We understand this effect in the following way. Schottky barriers form at the contacts to nanotube FETs[23,24] (Figure 2). Energy band pinning in such devices is commonly asymmetric, so holes conduct more readily than electrons.[26] Electron conduction is typically still measurable in large-diameter (small energy bandgap) SWNTs, leading to ambipolar $I(V_g)$ characteristics (Figure 3). In contrast, small diameter (large bandgap) SWNTs typically show p-type conduction, with electron conduction suppressed below measurement sensitivity (Figure 4). As described below, the data agree with a model where the Schottky barrier acts as a tunnel barrier with different, temperature-independent transparencies for the two carrier types. When $V_g$ is set so the Fermi energy $E_F$ lies in the band gap, transport occurs with an activation energy given by $E_a = |E_F - E_{band}|$, where $E_{band}$ is the energy band edge (valence or conduction) closest to $E_F$; the activation energy is therefore expected to vary linearly with $V_g$, reaching a maximum of half the energy band gap when $E_F$ is situated mid-gap.

We see precisely this behavior for the ambipolar sample Device I. Figure 3 shows the temperature dependence of $I(V_g)$ for this sample, the 4 nm diameter bundle imaged in Figure 1c. We used Scanning Impedance Microscopy to verify that this structure was the only current path connecting source and drain contacts. For fixed $V_g$, the source-drain current data show the expected thermally-activated dependence. We use an Arrhenius plot to extract an activation energy $E_a$ (Figure 4b, inset), which is plotted as a function of $V_g$ in Figure 3b.

The linear regions in Figure 3b (–2 V < $V_g$ < 2 V) occur when $E_F$ is situated in the band gap of the semiconducting SWNT. At $V_g = 0$, $E_a$ reaches a maximum of about 150 meV; the energy gap of this SWNT is therefore 300 meV, corresponding to a nanotube diameter near 2 nm that is compatible with AFM images of the structure (Figure 1c). From a linear fit to $E_a$ in the gap region, the ratio of gate capacitance to total capacitance, or "lever arm", is found to be $\alpha \approx 0.08$, similar to the value of 0.1 found for CVD-grown samples with the same device geometry.[8] The activation energy oscillates for –8 V < $V_g$ < –2 V as does $I(V_g)$. We attribute these oscillations to single electron charging and note that a maximum (minimum) in the activation energy near $V_g = -6$ V ($V_g = -8$ V, –4 V) corresponds to a minimum (maximum) in $I(V_g)$, as expected for the charging regime.

Figure 4a shows $I(V_g)$ data as a function of temperature for Device II, a p-type FET. Again we observe that $E_a$ increases linearly with gate voltage in the range –4 V < $V_g$ < 0 V, as expected within the model. We can not determine $E_a$ for positive $V_g$ because the current at low temperature is below measurement sensitivity, but the data indicate an energy gap greater than 400 meV and a lever arm $\alpha \approx 0.03$. Similar to Device I, $I(V_g)$ and $E_a(V_g)$ exhibit oscillations that are attributed to Coulomb effects.

In conclusion, we have demonstrated significant progress towards the goal of fabricating SWNT nanoeletronic devices from bulk HiPCO-grown material.

Devices fabricated from raw HiPCO have very high resistance; careful purification is thus essential for removing impurities that degrade device characteristics. After purification, resuspension, deposition, and surfactant removal, SWNTs retain the unique electronic properties that make them leading candidates for nanoelectronic devices. Finally, we demonstrate how the energy gap of individual semiconducting nanotubes can be quantitatively inferred from measurements of device current as a function of temperature and gate voltage.

## Methods

### Details of the purification process.

Wet air burn:
1. Impurity carbon phases (amorphous carbon, fullerenes, etc.) are removed by heating raw HiPCO material in air in the presence of $H_2O_2$ for 3–6 hours.

Acid treatment:
2. Oxidized SWNT material is refluxed with 2–3 M $HNO_3$ for 20 minutes, neutralized with NaOH, and then washed with deionized water.
3. Material is refluxed with $H_2O_2$ for 10 minutes.
4. Steps 2, 3 are repeated 2–3 times.

Annealing
5. Material is annealed in vacuum at 1150 °C for 2–3 hours.

Magnetic fractionation
6. SWNT material is dispersed in NaDDBS surfactant solution as detailed in Ref. 11. The material is flowed over a magnetic field gradient (~ 0.08 T/cm). Magnetic impurities feel a force due to the field gradient and are removed from the main flow of material.

Thermogravimetric analysis and wide-angle X-ray scattering measurements indicate impurity content is more than 50 wt% in as-grown HiPCO and less than 5 wt% after purification. Based on this measured impurity content and the measured sample mass after purification, the purification process recovers close to 90 wt% of the SWNT content of the raw HiPCO material.

## Acknowledgements

AGY acknowledges partial support from NSF DMR-0203378 and NASA (NAG8-2172). DEJ acknowledges support from NSF IGERT (DGE-0221664). AGY and ATJ acknowledge support from NSF MRSEC DMR-079909. ATJ acknowledges partial support from the Petroleum Research Fund. Correspondence and requests for materials should be addressed to MFI (islam@physics.upenn.edu) or ATJ (cjohnson@physics.upenn.edu).

## Competing financial interests

The authors declare that they have no competing financial interests.

## Supplementary Information

**Table 1. Observed Resistance Values**

|  | Run 1<br>Raw HiPCO | Run 2<br>Purified HiPCO | Run 3<br>Purified HiPCO |
|---|---|---|---|
| Device Yield | 4/16 | 7/30 | 29/98 |
| SC Devices<br>Resistances (MΩ) | 2<br>700, 2000 | 1<br>140 | 7<br>1.4, 10, 10, 10, 20, 250, 1400 |
| M/H Devices<br>Resistances (MΩ) | 2<br>570, 1300 | 6<br>0.5, 0.5, 4, 8, 10, 2000 | 22<br>Median = 1<br>Mean = 0.5 |

Run 1 and Run2 use material from the same batch of HiPCO.
Measured resistance values (MΩ) for Run 3, M/H devices are: 0.1, 0.13, 0.15 (x2), 0.17 (x3), 0.25 (x2), 0.3, 0.5 (x2), 0.55, 0.6, 1, 1.4, 2, 3 (x2), 5.